\documentclass{article}
\usepackage{slt/spconf,amsmath,graphicx}
\usepackage{multirow}
\usepackage{booktabs}
\usepackage{hyperref}
\usepackage{url}

\title{SOCODEC: A SEMANTIC-ORDERED MULTI-STREAM SPEECH CODEC FOR \\ EFFICIENT LANGUAGE MODEL BASED TEXT-TO-SPEECH SYNTHESIS}
\name{BLIND}
\address{BLIND}

\name{Haohan Guo$^*$,
    \thanks{Work performed during the first author's internship at Xiaohongshu. This work is supported by the Centre for Perceptual and Interactive Intelligence (CPII) Ltd under the Innovation and Technology Fund.}
    Fenglong Xie$^\dag$,
    Kun Xie$^\dag$,
    Dongchao Yang$^*$,
    Dake Guo$^\ddag$,,
    Xixin Wu$^*$,
    Helen Meng$^*$}
\address{
    $^*$The Chinese University of Hong Kong, Hong Kong SAR, China \\
    $^\dag$Xiaohongshu Inc., Shanghai, China \\
    $^\ddag$Northwestern Polytechnical University, Xi’an, China \\
    \href{mailto:hguo@se.cuhk.edu.hk}{\nolinkurl{{hguo, dcyang, xxwu, hmmeng}@se.cuhk.edu.hk}}, \\
    \href{mailto:fenglongxie@xiaohongshu.com}{{\nolinkurl{{fenglongxie, weisi}@xiaohongshu.com}}}, 
    \href{mailto:guodake@mail.nwpu.edu.cn}{\nolinkurl{guodake@mail.nwpu.edu.cn}}
}

\begin{document}
%
\maketitle
\begin{abstract}

The long speech sequence has been troubling language models (LM) based TTS approaches in terms of modeling complexity and efficiency. This work proposes SoCodec, a semantic-ordered multi-stream speech codec, to address this issue. It compresses speech into a shorter, multi-stream discrete semantic sequence with multiple tokens at each frame. Meanwhile, the ordered product quantization is proposed to constrain this sequence into an ordered representation. It can be applied with a multi-stream delayed LM to achieve better autoregressive generation along both time and stream axes in TTS. The experimental result strongly demonstrates the effectiveness of the proposed approach, achieving superior performance over baseline systems even if compressing the frameshift of speech from 20ms to 240ms (12x). The ablation studies further validate the importance of learning the proposed ordered multi-stream semantic representation in pursuing shorter speech sequences for efficient LM-based TTS.

\end{abstract}

\section{Introduction}

Large language models (LLMs) have demonstrated powerful capability in text generation \cite{brown2020language,openai2023gpt4,touvron2023llama2}. This breakthrough inspires applications of LLMs in speech generation, especially for zero-shot text-to-speech (TTS) synthesis, e.g., VALL-E \cite{VALLEX}, Tortoise \cite{tortoise}, BASE-TTS \cite{lajszczak2024base}, etc. These models treat TTS as a next-token prediction task by auto-regressively generating discrete speech tokens (codes). Hence, such a language-model-based TTS system, i.e. LM-TTS, usually relies on an audio codec system \cite{encodec, hifi-codec, dac}. It encodes speech signals into discrete speech tokens and reconstructs them back. Then, we train an LM to predict speech tokens from the text to achieve TTS synthesis. However, unlike the text, speech signals contain abundant information, including phonetics, prosody, speaker identity, acoustic environment, etc., making it challenging to compress into a token sequence as short as the text. The long sequence seriously affects the LM in terms of modeling complexity and efficiency, hindering its development in the speech domain.

In this work, we propose SoCodec, a semantic-ordered multi-stream speech codec, to provide a shorter token sequence for efficient LM-TTS. First, SoCodec leverages the self-supervised-learning-based model \cite{hsu2021hubert, wavlm} to compress speech signals into a multi-stream semantic sequence containing sufficient phonetic and prosodic information. Then, an utterance-level acoustic embedding is extracted from the Mel spectrogram to represent time-invariant acoustic information, including speaker identity, acoustic environment, etc. Meanwhile, we propose ordered product quantization (OPQ) for SoCodec to quantize speech into an ordered speech representation along the stream axis. It can be incorporated with a multi-stream LLM \cite{copet2024simple} based on a delayed pattern to achieve the high-quality and efficient zero-shot TTS, which is validated in both subjective and objective experiments.

Our contributions are summarized as follows: 1) we propose a new speech codec, SoCodec, providing a shorter speech sequence for efficient LM-TTS; 2) we propose ordered product quantization (OPQ) to learn an ordered multi-stream sequence to adapt multi-stream LM better; 3) we propose an LM-TTS system based on SoCodec, achieving higher efficiency while keeping high synthesis quality in TTS, even with a frameshift of only 240ms, the shortest sequence across all LM-TTS approaches to the best of our knowledge.

\section{Related Work}

To provide a short speech sequence for LMs, the codec is usually optimized from two aspects: 1) information reduction, compressing speech signals to represent speech with fewer tokens, and 2) representing speech with multiple streams, i.e., each frame consists of multiple tokens to increase the frameshift of the sequence. The mainstream audio codec, e.g. Encodec \cite{encodec}, Hifi-Codec \cite{hifi-codec}, DAC \cite{hifi-codec}, usually adopt residual vector quantization (RQ) based approaches to compress speech signals into a multi-stream sequence with a frameshift of around 20ms and more than 8 streams to cover sufficient acoustic information. However, the sequence is still too long to adapt LMs well. Hence, TiCodec \cite{ren2024fewer} and SingleCodec \cite{li2024single} are proposed to represent speech with fewer tokens by disentangling time-invariant acoustic information out of the discrete sequence. Meanwhile, some works \cite{ns3, zhang2023speechtokenizer, liu2024semanticodec} emphasize keeping only semantic information in speech tokens to compress the sequence further. Based on these works, we propose SoCodec to compress speech into a shorter multi-stream semantic representation.

On the other hand, to better generate the multi-stream representation, an ordered generation process along the stream axis is also applied in LM-TTS. For example, VALL-E \cite{wang2023neural, VALLEX} predicts the first stream via one AR model, and predicts the following streams recursively by running one NAR model 7 times. Recently, a multi-stream LLM with a delayed pattern \cite{copet2024simple, kharitonov2022text} is proposed to generate tokens along both axes auto-regressively by only running one AR model once. However, this autoregressive generation from the low stream to the high stream makes it easy to deliver accumulated errors to high-stream prediction, degrading the generation quality. Hence, we intuitively aim to seek an ordered multi-stream speech representation to first generate principal speech information in low streams to ensure a stable generation process. Inspired by ordered representation learning \cite{rippel2014learning, xu2021anytime}, we propose ordered product quantization for SoCodec.

\section{Semantic-Ordered Speech Codec}

In this section, we will give a detailed introduction to SoCodec, including the proposed ordered product quantization, the model architecture, and the loss function.

\subsection{Ordered Product Quantization}

Ordered representation learning \cite{rippel2014learning, xu2021anytime} aims to encode the input into a PCA-like representation, where the first dimension represents the most principal information of the input, and the following dimensions represent the residual information recursively. Inspired by it, we propose ordered product quantization (OPQ) to encode speech into a representation with this order along the stream axis.

\begin{figure}[t]
    \centering
    \includegraphics[width=0.8\linewidth]{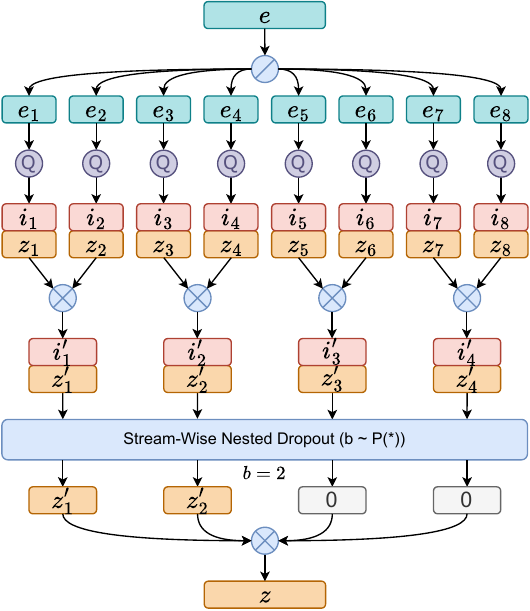}
    \caption{An example of the four-stream ordered product quantization. $\bigotimes$ denotes the concatenation operation.}
    \label{fig:opq}
\end{figure}

Fig. \ref{fig:opq} shows how a vector $e$ is encoded to a four-stream representation by OPQ in training, consisting of two parts: grouped product quantization, and stream-wise nested dropout. First, like the vanilla product quantization \cite{jegou2010product}, we chunk $e$ into 8 sub-vectors with the same dimension, and quantize them respectively with different codebooks ${c_1, c_2, ..., c_8}$. Each input sub-vector $e_i$ is converted to a quantized sub-vector and the index to its corresponding codebook. Then, following \cite{guo2024addressing}, we combine each two quantized sub-vectors to form one stream, i.e. a concatenated sub-vector, $z^{'}_1 = [z_1, z_2]$, and a new index, $i^{'}_1 = i_1 * |c_2| + i_2$, where $|c_2|$ is the codebook size of $c_2$. In this way, we can easily learn large-codebook ($|c^{'}_1| = |c_1| * |c_2|$) representations for LMs while avoiding codebook collapse. Then, to train these streams to be ordered, we follow the ordered autoencoder \cite{rippel2014learning}, and propose a stream-wise nested dropout. Given a random number $b$ in training, we keep only the first $b$ streams and mask the rest streams with 0 to form the output vector $z$. In this way, we explicitly train SoCodec to encode speech into an ordered multi-stream representation. In this work, we sample $b$ from a uniform distribution. Notably, nested dropout is disabled at inference to provide the complete ordered representation.

\subsection{Model Architecture}


\subsubsection{Semantic Encoder}

\begin{figure}[t]
    \centering
    \includegraphics[width=0.8\linewidth]{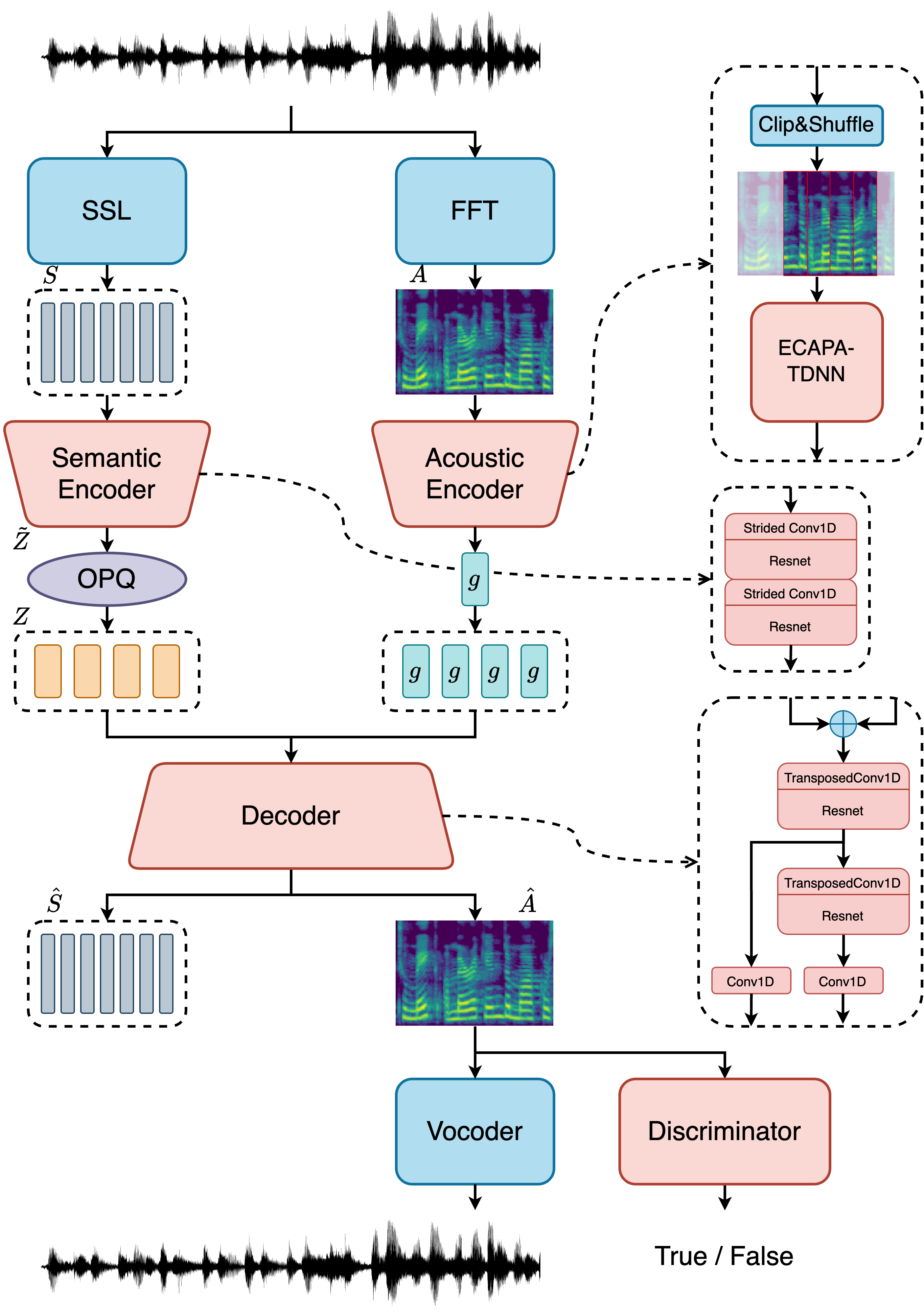}
    \caption{The model architecture of SoCodec.}
    \label{fig:socodec}
\end{figure}

As shown in Fig. \ref{fig:socodec}, we first encode the speech signal into a semantic token sequence by employing a pre-trained self-supervised learning (SSL) model, HuBERT \cite{hsu2021hubert}. It encodes speech signals into embedding sequence $S$ with rich semantic information as the encoder input. The time-variant encoder based on ResNet blocks further processes this sequence and uses stridden convolutional layers for down-sampling to obtain a shorter encoding sequence. We then process this sequence with OPQ to obtain the ordered quantized sequence $Z$.

\subsubsection{Acoustic Encoder}

Meanwhile, we also apply an ECAPA-TDNN-based \cite{dawalatabad2021ecapa} time-invariant encoder to extract an utterance-level global embedding $g$ from speech signals to preserve time-invariant information, e.g., speaker identity, global speaking style, acoustic environment, etc. This embedding can be used in speech reconstruction and imitate the target voice in zero-shot TTS. To avoid the leakage of content information into this embedding, we propose a simple pro-processing technique, ``Clip\&Shuffle", on the Mel spectrogram to remove short-time variant information from it. Specifically, we first sample a segment with a length of 25\% to 75\% of the utterance, and then chunk it into slices with the length of 1 second. Finally, these slices are shuffled randomly to form a new sequence. This simple approach effectively reduces content leakage, and avoids complicated disentangling techniques such as adversarial learning \cite{lajszczak2024base}.

\subsubsection{Decoder}

In the decoder, the global embedding $g$ is duplicated and added with the quantized sequence $Z$ to form the decoder input. It is then processed by ResNet blocks with the transposed convolutional layers for up-sampling to reconstruct both SSL features and acoustic features. The discriminator proposed in Mega-TTS \cite{jiang2023mega} is applied in training to improve the generative quality of the Mel spectrogram. Finally. we employ a pre-trained Mel-spectrogram-based neural vocoder, BigVGAN \cite{bigvgan}, to generate the reconstructed audio. Notably, although the SSL feature is not applied to generate the audio, the training objective of minimizing the reconstruction loss of SSL features still matters in learning discrete semantic representations.

\subsubsection{Loss Function}

The loss function of SoCodec is written as follows:
\begin{align}
    \mathcal{L}_c &= \lambda_1 * \left \| Z - \tilde{Z} \right \|^2_2 + \lambda_2 * \left \| S - \hat{S} \right \|^2_2 \\
    & + \lambda_3 * \left \| A - \hat{A} \right \|^2_2 + \lambda_4 * \left \| 1 - D(\hat{A}) \right \|^2_2
\end{align}
where $\lambda_1, \lambda_2, \lambda_3, \lambda_4$ are weight coefficients. $\left \| Z - \tilde{Z} \right \|^2_2$ is the VQ loss between the quantized sequence $Z$ and the encoding sequence $\tilde{Z}$ before the quantization. $\left \| S - \hat{S} \right \|^2_2$ is the semantic loss between the SSL features $S$ and the reconstructed ones $\hat{S}$. Finally, the acoustic loss is composed of two terms: the L2 loss between the ground-truth Mel spectrogram $A$ and the reconstructed one $\hat{A}$, and an adversarial loss, where $D(*)$ denotes outputs of all discriminators. Meanwhile, discriminators are trained alternately with codec.

\section{Multi-Stream Language Model}

\subsection{Model Architecture}

\begin{figure}[htp]
    \centering
    \includegraphics[width=0.9\linewidth]{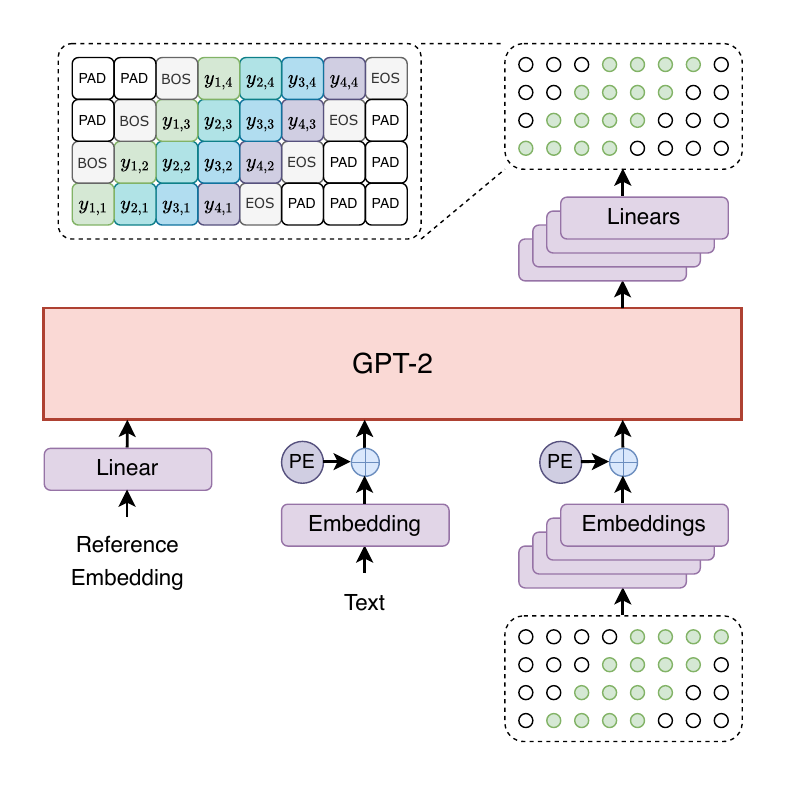}
    \caption{The model structure of the multi-stream LM. ``PE'' denotes the learnable positional embedding layer.}
    \label{fig:lm}
\end{figure}

In this work, we adopt a GPT-2-based \cite{radford2019language} decoder-only Transformer to construct a multi-stream LM, as shown in Fig. \ref{fig:lm}. In training, we first extract semantic tokens and global (reference) embedding from the target speech via the pre-trained SoCodec. Then, the reference embedding, text, and speech, are mapped into the embedding space with the same dimension. All streams of the speech sequence are processed respectively and then added together. The text sequence and speech sequence are added with different learnable positional embeddings. Then, we process this embedding sequence with a stack of causal Transformer layers, and predict probabilities of speech tokens with a group of linear layers. The loss function of multi-stream LM is the averaged cross-entropy loss across all streams of the speech sequence.

\subsection{Delay Prediction}

An ideal LM for multi-stream representations is to predict the joint distribution of all streams in an auto-regressive manner:
\begin{align}
    P(y_{t, 1:m}|y_{1: t-1, 1: m}, x)
\end{align}
where $m$ and $x$ are the number of streams and the input, i.e. the text and the reference embedding in our work. It requires integrating all tokens at the same frame into one token corresponding to a giant dictionary, which is impossible to achieve. Hence, we usually adopt the chain rule, i.e.:
\begin{align}
    \prod^m_{j=1}P(y_{t, j}|y_{t, 1: j-1}, y_{1: t-1, 1: m}, x)
\end{align}
we can achieve this by flattening all streams into one sequence, but it also multiplies the length of the sequence, going against our intention of using multi-stream representations. The delayed prediction breaks the dilemma effectively and has shown superior performance in music and speech generation \cite{copet2024simple, kharitonov2022text}. It shifts the $j$-th stream with $d * (j-1)$ frames for all streams, where $d$ is the pre-designed number of delayed steps. In this way, the objective of LM is changed to:
\begin{equation}
\begin{aligned}
\label{eq:delay}
    \prod^m_{j=1}P(y_{t, j}|&y_{1: t - 1 + d*(j - 1), 1}, y_{1: t - 1 + d*(j - 2), 2}, \\
    & ..., y_{1: t - 1 + d*(j - m), m}, x)
\end{aligned}
\end{equation}
This approach allows us to predict multiple tokens in parallel while keeping chain-rule prediction from low to high streams. However, this pattern also causes missing high-stream information in low-stream prediction and more accumulated errors in high-stream prediction. Specifically, when generating low-stream tokens, the model cannot see full information on high streams from the past. For example, $y_{3,4}$ is unseen when generating $y_{4, 1}$. Moreover, the accumulated errors from the auto-regressive prediction lead to more noise in the high-stream prediction. Hence, an ordered representation is necessary for this multi-stream LM to ensure that principal speech information is preserved and predicted in low streams for a high-quality and stable generation.


\section{Experimental Protocol}

\subsection{Datasets \& Features}

All codecs and LLMs in this work are trained with WenetSpeech4TTS (Basic) \cite{ma2024wenetspeech4tts}, an open-source large-scale Chinese multi-speaker speech dataset. It contains 7,226 hours of found data with high diversity in audio quality, speaking style, acoustic environment, etc., hence putting a huge challenge to TTS modeling. All audio files are normalized to the sample rate of 16kHz in our experiments.  SoCodec adopts a pre-trained Chinese HuBERT \cite{TencentGameMate} to extract 1024-dim SSL features with a frameshift of 20ms, and uses the 80-dim Mel-spectrogram with a frameshift of 10ms as the acoustic feature. The BigVGAN-base \cite{bigvgan} neural vocoder is pre-trained on WenetSpeech4TTS to synthesize audio from Mel spectrograms. We train all LMs to generate speech tokens from the normalized text directly. Hence, we train a byte-pair encoding (BPE) based text dictionary with 8192 tokens, and feed text tokens to our LMs.

\subsection{Models}
\label{ssec:model}

In SoCodec\footnote{The implementation is available at \url{https://github.com/hhguo/SoCodec}}, the time-variant encoder and the decoder are applied with 1024-dim ResNet blocks, each consisting of 4 convolutional layers. We apply the time-invariant encoder with a 256-dim ECAPA-TDNN based on a smaller ResNet block with only 2 convolutional layers. In OPQ, we fix the number of codewords of each codebook to 128, and set the dimension of input vectors of the OPQ module to the total number of codewords, e.g. 1024 for 8 codebooks. Two codebooks form a larger codebook with 16,384 codewords for LLM training and inference. In training, we set $\lambda_1=1, \lambda_2=10^3, \lambda_3=10, \lambda_4=1$ to balance these loss terms, and use EMA to update codebooks with the decay rate of 0.99.

LMs use the GPT-2 module consisting of 12 Transformer layers with a feature dimension of 1024 to predict probability distributions with the dimension of 16386 (the extra two indices are for BOS and EOS tokens) for each stream. SoCodec and LMs are trained using AdamW \cite{loshchilov2017decoupled} for 100,000 iterations. By default, we train models with a batch size of 1600 seconds and run LM inference using a sampling strategy with a temperature of 0.8, a top-p of 0.8, a top-k of 10, and a repetition penalty of 2.0 for a stable generation process. In \ref{ssec:tts_comparison}, we increase the batch size to 3.5 hours in training to keep a similar training configuration with baseline systems.

\subsection{Evaluation Metrics}

In codec evaluation, we sample 860 utterances from the WenetSpeech \cite{zhang2022wenetspeech} test set, with high diversity in speaking style and audio quality, to evaluate the performance of codecs. We adopt the following objective metrics\footnote{We use FunASR, the open-source ASR tool, for transcribing, available at \url{https://github.com/modelscope/FunASR}. The tool for extracting speaker embedding is available at \url{https://huggingface.co/Wespeaker/wespeaker-cnceleb-resnet34}}: Mel-cepstrum distortion (MCD, dB), character error rate (CER, \%), and speaker similarity (SIM, $\times10^{-2}$) to measure reconstruction quality, intelligibility, and speaker similarity. 

In TTS evaluation, we create a test set with 860 utterances, where each utterance is paired with a different out-of-training-set audio file as the reference audio for zero-shot TTS. In \ref{ssec:tts_comparison}, we use the subset with 100 utterances to conduct objective and subjective tests, i.e. the MOS tests. There are 10 native speakers invited to the test and asked to give scores ranging from 1 to 5 in terms of naturalness (NMOS) and speaker/style similarity (SMOS), respectively. We also calculate the real-time factor (RTF) to measure the efficiency of TTS systems.

\begin{table}[htp]
\centering
\begin{tabular}{l|ccccc}
\toprule
             & NMOS & SMOS & CER & SIM & RTF \\ \midrule
X-TTS        & 3.51 & 2.83 & 4.39  & 76.89 & 0.47 \\
VALL-E       & 3.40 & 2.73 & 10.69 & 72.45 & 0.95 \\
SoCodec-40   & 3.83 & \textbf{3.91} & \textbf{2.37}  & \textbf{85.17} & 0.46 \\
SoCodec-120  & \textbf{3.98} & 3.78 & 2.57  & 83.14 & 0.22 \\ 
SoCodec-240  & 3.77 & 3.47 & 3.01  & 79.02 & \textbf{0.16} \\ \bottomrule
\end{tabular}
\caption{The comparison of different LM-TTS systems.}
\label{tab:mos}
\end{table}

\section{Results}

\subsection{TTS System Comparison}
\label{ssec:tts_comparison}

First, we use both subjective and objective metrics to compare different TTS systems\footnote{X-TTS is available at \url{https://huggingface.co/coqui/XTTS-v2}, and VALL-E is available at \url{https://github.com/dukGuo/valle-audiodec}}: X-TTS, an industrial zero-short baseline LM-TTS system; VALL-E trained on the same datasets; and SoCodec-based TTS systems. We propose three versions of SoCodec-based TTS: SoCodec-40 is a single-stream codec with a frameshift of 40ms; SoCodec-120 is a four-stream codec with a frameshift of 120ms; and SoCodec-240 is an eight-stream codec with a frameshift of 240ms.

As shown in Fig. \ref{tab:mos}, the baseline system VALL-E performs poorly on this challenging dataset, showing a worse quality than the industrial baseline X-TTS. It is based on a general audio codec with eight streams and a short frameshift of 20ms. The long speech sequence and the complicated framework (one AR inference and seven NAR inferences) cause high modeling complexity and low efficiency with only an RTF of 0.95. After emphasizing encoding semantic information in the codec, the single-stream LM-TTS, SoCodec-40, achieves the best overall quality across all models. By providing the proposed ordered multi-stream representation, we shorten the sequence by three times in SoCodec-120, leading to a higher efficiency with the RTF of 0.22, while keeping a comparable TTS quality to SoCodec-40. Finally, we try to further shorten the sequence to the frameshift of 240ms. It shows a slight degradation in TTS quality but the highest efficiency, and still significantly outperforms VALL-E based on the sequence 12 times longer. It strongly demonstrates the effectiveness of the proposed approach in achieving a high-quality and efficient LM-TTS.\footnote{Samples are available at \url{https://hhguo.github.io/DemoSoCodec}}

\begin{table}[htp]
\centering
\caption{The objective evaluation of speech audio from analysis-synthesis and TTS synthesis}
\label{tab:codec}
\begin{tabular}{c|ccc|cc}
\toprule
\multirow{2}{*}{System} & \multicolumn{3}{c|}{Analysis-Synthesis} & \multicolumn{2}{c}{TTS} \\ \cmidrule{2-6}
        & MCD & CER & SIM & CER & SIM \\ \midrule
Codec-1 & 6.38 & 24.94 & \textbf{84.64} & 14.71 & 81.03 \\
Codec-2 & 6.36 & 12.47 & 81.78 & 3.86 & 79.22 \\
SoCodec & \textbf{6.17} & \textbf{6.41} & 81.75 & \textbf{2.60} & \textbf{81.12} \\ \bottomrule
\end{tabular}
\end{table}

\subsection{Semantic Codec}

To evaluate the effect of semantic encoding in LM-TTS, we compare SoCodec with two codecs: Codec-1, replacing HuBERT with the Mel spectrogram and removing semantic loss, and Codec-2, only removing the semantic loss of SoCodec. These codecs are all based on four-stream speech sequences with a frameshift of 120ms. As shown in Table \ref{tab:codec}, Codec-1 preserves rich acoustic information, showing the highest speaker similarity, but also loses much semantic information, leading to the highest CER in analysis-synthesis and TTS. After adopting the HuBERT feature as the input, Codec-2 preserves more semantic information in the token sequence, improving the intelligibility of the reconstructed and TTS-synthesized audio with slightly degraded speaker similarity. Finally, the SoCodec trained with semantic loss further improves intelligibility while keeping a high speaker similarity. This result demonstrates that the proposed approach encodes sufficient semantic information to the discrete sequence to better produce intelligible speech in TTS.

\begin{figure}[htp]
    \centering
    \includegraphics[width=\linewidth]{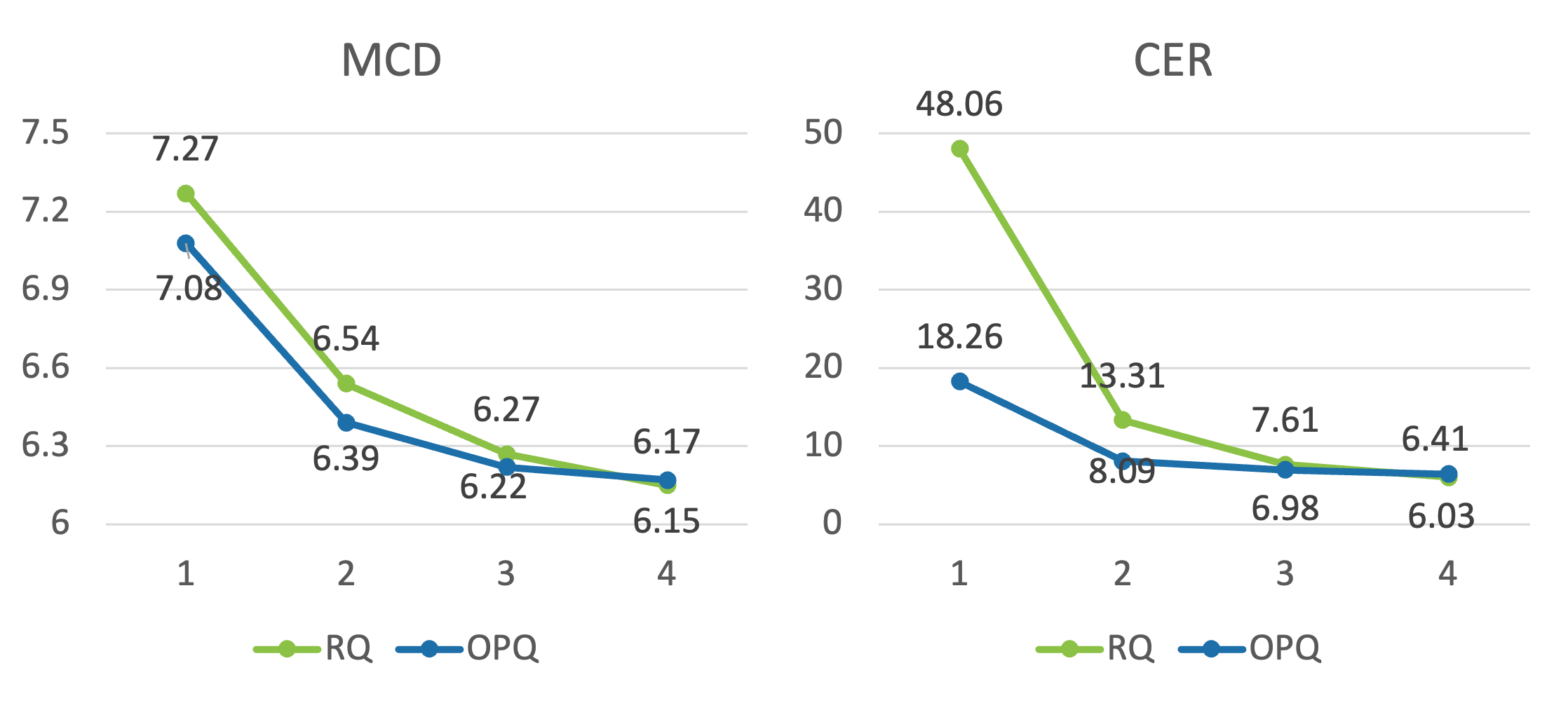}
    \caption{The MCD and CER of speech reconstructed from first $k$ streams in RQ and OPQ-based codecs.}
    \label{fig:order}
\end{figure}

\subsection{Ordered Speech Representations}

To investigate the impact of OPQ on TTS, we first make an analysis to see if OPQ can provide the expected ordered representation. Fig. \ref{fig:order} shows curves of MCD and CER of speech reconstructed from the first $b$ streams in RQ-based and OPQ-based four-stream SoCodec with a frameshift of 120ms. RQ can also provide an ordered representation approximately by quantizing the vector recursively in residual spaces. However, OPQ is shown as a more significant ordered representation. These two approaches achieve similar reconstruction quality when all streams are used, but OPQ preserves more principal information in lower streams, showing lower MCD and CER. It demonstrates the effectiveness of OPQ in learning the expected ordered multi-stream representation.

\begin{figure}[htp]
    \centering
    \includegraphics[width=\linewidth]{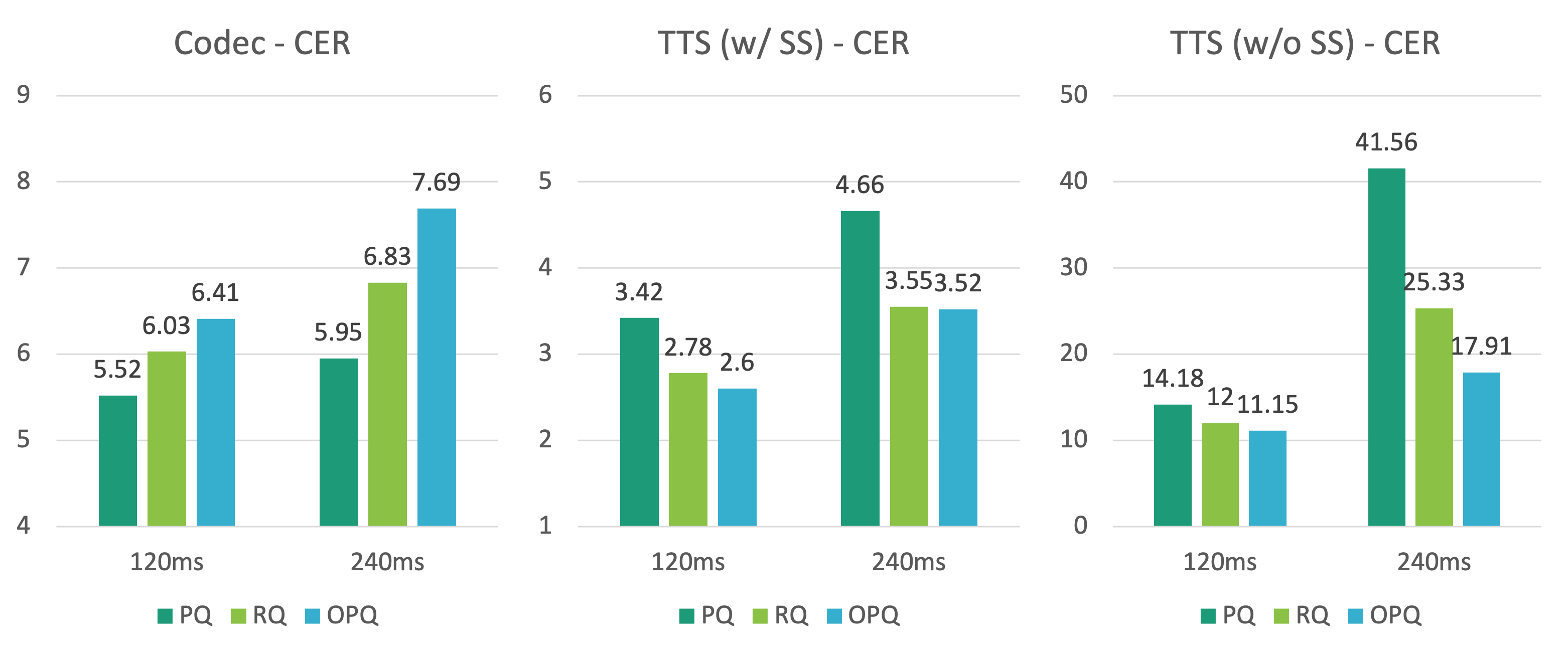}
    \caption{The CERs of audio from analysis-synthesis, TTS with the sampling strategy, and TTS without the sampling strategy, using SoCodec with various VQ approaches and frameshifts.}
    \label{fig:vq}
\end{figure}

Then, we investigate the impact of an ordered speech representation on analysis-synthesis and TTS by comparing SoCodec based on different VQ approaches, PQ, RQ, and OPQ, and different frameshifts, 120ms, and 240ms, as shown in Fig. \ref{fig:vq}. First, PQ-based SoCodec, without the ordered constraint, can fully utilize the embedding space to minimize the reconstruction loss, showing the lowest CER. RQ and OPQ bring more reconstruction loss to keep the expected order along the stream axis. However, ordered speech sequences benefit TTS significantly. We run TTS in two modes: with (TTS w/ SS) or without (TTS w/o SS) the sampling strategy mentioned in \ref{ssec:model} to evaluate the robustness of the model to noisy samples. The result shows that both RQ and OPQ improve TTS quality over the PQ-based system, especially on the longer frameshift of 240ms, but the OPQ shows the best performance in both TTS modes by providing a better ordered representation. Moreover, OPQ with the lowest CER in TTS w/o SS further validates that it makes the multi-stream autoregressive generation more robust to accumulated errors.

\subsection{Delayed Prediction}

\begin{figure}[htp]
    \centering
    \includegraphics[width=0.75\linewidth]{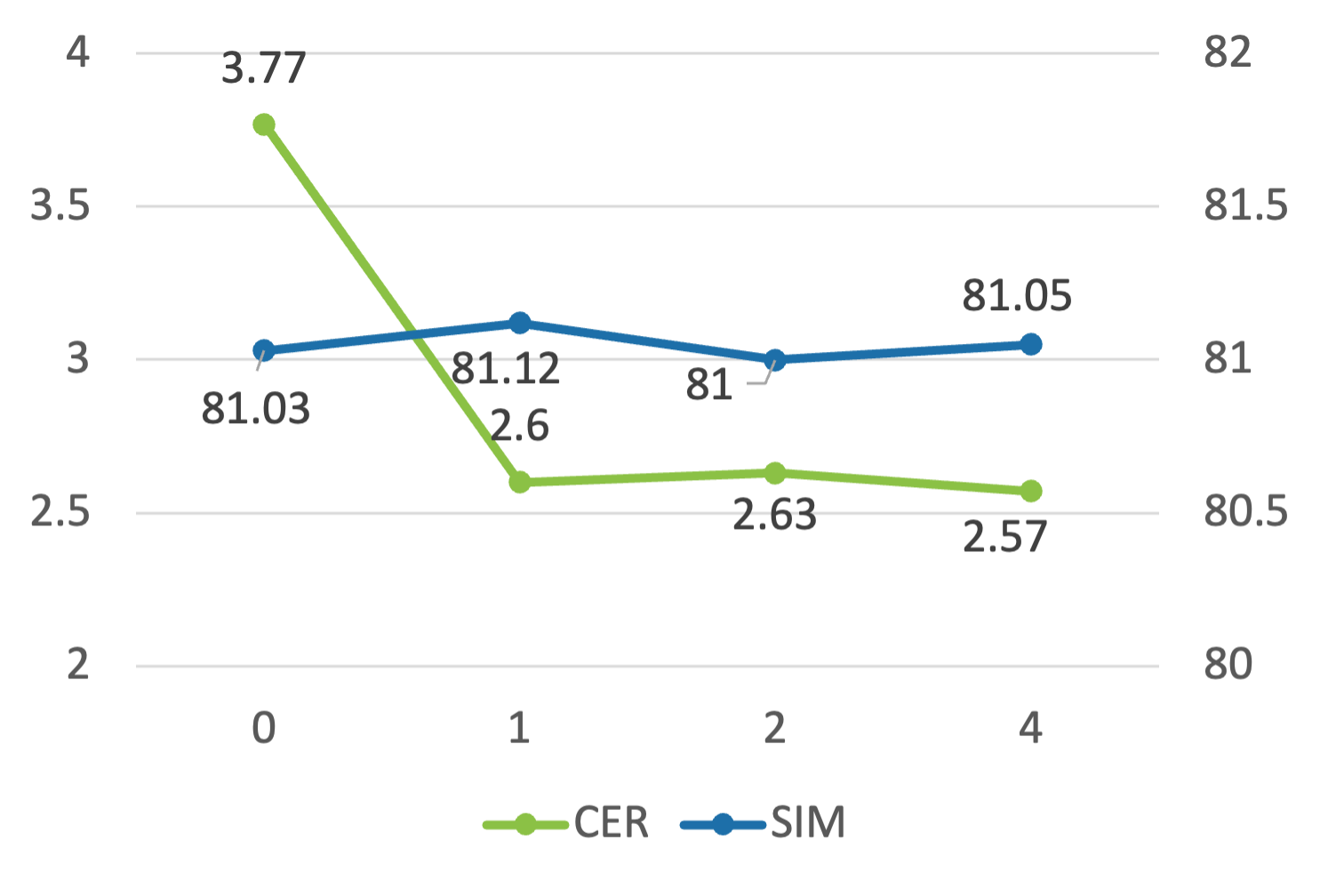}
    \caption{The CERs of synthesized audio from LLMs with different delayed steps.}
    \label{fig:delay}
\end{figure}

We also investigate the impact of delayed prediction of the multi-stream LLM on TTS quality. Fig. \ref{fig:delay} shows CERs and SIMs of SoCodec-120 with different delayed steps. First, LMs with different delayed steps show similar performance in speaker similarity. However, the LM without delayed prediction ($d=0$) produces more unintelligible and unnatural speech, showing the highest CER. It verifies the necessity of the delayed pattern in multi-stream LM. Moreover, more delayed steps bring more computing costs but no further significant improvement. We notice that, as indicated in Eq. \ref{eq:delay}, more delayed steps make high-stream prediction receive more information from low streams but also make low-stream prediction lose more information from high streams. This trade-off makes it harder to gain from more delayed steps. Hence, we conclude that combining SoCodec with multi-stream LM with the delayed step of 1 can already achieve a high-quality and efficient LM-TTS. 

\section{Conclusions}

We propose a new speech codec, SoCodec, to provide a shorter multi-stream speech sequence for efficient LM-TTS. Meanwhile, ordered product quantization (OPQ) is proposed to learn an ordered multi-stream sequence to be better incorporated with the multi-stream delayed LM to achieve high-quality and efficient zero-shot TTS. Finally, the proposed LM-TTS system based on SoCodec achieves outperforming TTS quality over baseline systems, while keeping much higher efficiency with shorter speech sequences. In ablation studies, SoCodec is validated as an effective model in encoding sufficient semantic information into the discrete sequence. The proposed ordered product quantization shows its significance in providing the expected ordered multi-stream representation, improving the performance of the multi-stream delayed LM in TTS.

\bibliographystyle{slt/IEEEbib}
\bibliography{refs}

\begin{thebibliography}{10}

\bibitem{brown2020language}
Tom Brown, Benjamin Mann, Nick Ryder, Melanie Subbiah, Jared~D Kaplan, Prafulla Dhariwal, Arvind Neelakantan, Pranav Shyam, Girish Sastry, Amanda Askell, et~al.,
\newblock ``Language models are few-shot learners,''
\newblock {\em Advances in neural information processing systems}, vol. 33, pp. 1877--1901, 2020.

\bibitem{openai2023gpt4}
OpenAI,
\newblock ``Gpt-4 technical report,'' 2023.

\bibitem{touvron2023llama2}
Hugo Touvron, Louis Martin, Kevin Stone, Peter Albert, Amjad Almahairi, Yasmine Babaei, Nikolay Bashlykov, Soumya Batra, Prajjwal Bhargava, Shruti Bhosale, et~al.,
\newblock ``Llama 2: Open foundation and fine-tuned chat models,''
\newblock {\em arXiv preprint arXiv:2307.09288}, 2023.

\bibitem{VALLEX}
Ziqiang Zhang, Long Zhou, Chengyi Wang, Sanyuan Chen, Yu~Wu, Shujie Liu, Zhuo Chen, Yanqing Liu, Huaming Wang, Jinyu Li, Lei He, Sheng Zhao, and Furu Wei,
\newblock ``Speak foreign languages with your own voice: Cross-lingual neural codec language modeling,''
\newblock {\em CoRR}, vol. abs/2303.03926, 2023.

\bibitem{tortoise}
James Betker,
\newblock ``Better speech synthesis through scaling,''
\newblock {\em arXiv preprint arXiv:2305.07243}, 2023.

\bibitem{lajszczak2024base}
Mateusz {\L}ajszczak, Guillermo C{\'a}mbara, Yang Li, Fatih Beyhan, Arent van Korlaar, Fan Yang, Arnaud Joly, {\'A}lvaro Mart{\'\i}n-Cortinas, Ammar Abbas, Adam Michalski, et~al.,
\newblock ``Base tts: Lessons from building a billion-parameter text-to-speech model on 100k hours of data,''
\newblock {\em arXiv preprint arXiv:2402.08093}, 2024.

\bibitem{encodec}
Alexandre D{\'e}fossez, Jade Copet, Gabriel Synnaeve, and Yossi Adi,
\newblock ``High fidelity neural audio compression,''
\newblock {\em arXiv preprint arXiv:2210.13438}, 2022.

\bibitem{hifi-codec}
Dongchao Yang, Songxiang Liu, Rongjie Huang, Jinchuan Tian, Chao Weng, and Yuexian Zou,
\newblock ``Hifi-codec: Group-residual vector quantization for high fidelity audio codec,''
\newblock {\em arXiv preprint arXiv:2305.02765}, 2023.

\bibitem{dac}
Rithesh Kumar, Prem Seetharaman, Alejandro Luebs, Ishaan Kumar, and Kundan Kumar,
\newblock ``High-fidelity audio compression with improved rvqgan,''
\newblock {\em Advances in Neural Information Processing Systems}, vol. 36, 2024.

\bibitem{hsu2021hubert}
Wei-Ning Hsu, Benjamin Bolte, Yao-Hung~Hubert Tsai, Kushal Lakhotia, Ruslan Salakhutdinov, and Abdelrahman Mohamed,
\newblock ``Hubert: Self-supervised speech representation learning by masked prediction of hidden units,''
\newblock {\em IEEE/ACM Transactions on Audio, Speech, and Language Processing}, vol. 29, pp. 3451--3460, 2021.

\bibitem{wavlm}
Sanyuan Chen, Chengyi Wang, Zhengyang Chen, Yu~Wu, Shujie Liu, Zhuo Chen, Jinyu Li, Naoyuki Kanda, Takuya Yoshioka, Xiong Xiao, Jian Wu, Long Zhou, Shuo Ren, Yanmin Qian, Yao Qian, Michael Zeng, Xiangzhan Yu, and Furu Wei,
\newblock ``Wavlm: Large-scale self-supervised pre-training for full stack speech processing,''
\newblock {\em IEEE Journal of Selected Topics in Signal Processing}, vol. 16, pp. 1--14, 10 2022.

\bibitem{copet2024simple}
Jade Copet, Felix Kreuk, Itai Gat, Tal Remez, David Kant, Gabriel Synnaeve, Yossi Adi, and Alexandre D{\'e}fossez,
\newblock ``Simple and controllable music generation,''
\newblock {\em Advances in Neural Information Processing Systems}, vol. 36, 2024.

\bibitem{ren2024fewer}
Yong Ren, Tao Wang, Jiangyan Yi, Le~Xu, Jianhua Tao, Chu~Yuan Zhang, and Junzuo Zhou,
\newblock ``Fewer-token neural speech codec with time-invariant codes,''
\newblock in {\em ICASSP 2024-2024 IEEE International Conference on Acoustics, Speech and Signal Processing (ICASSP)}. IEEE, 2024, pp. 12737--12741.

\bibitem{li2024single}
Hanzhao Li, Liumeng Xue, Haohan Guo, Xinfa Zhu, Yuanjun Lv, Lei Xie, Yunlin Chen, Hao Yin, and Zhifei Li,
\newblock ``Single-codec: Single-codebook speech codec towards high-performance speech generation,''
\newblock {\em arXiv preprint arXiv:2406.07422}, 2024.

\bibitem{ns3}
Zeqian Ju, Yuancheng Wang, Kai Shen, Xu~Tan, Detai Xin, Dongchao Yang, Yanqing Liu, Yichong Leng, Kaitao Song, Siliang Tang, et~al.,
\newblock ``Naturalspeech 3: Zero-shot speech synthesis with factorized codec and diffusion models,''
\newblock {\em arXiv preprint arXiv:2403.03100}, 2024.

\bibitem{zhang2023speechtokenizer}
Xin Zhang, Dong Zhang, Shimin Li, Yaqian Zhou, and Xipeng Qiu,
\newblock ``Speechtokenizer: Unified speech tokenizer for speech large language models,''
\newblock {\em arXiv preprint arXiv:2308.16692}, 2023.

\bibitem{liu2024semanticodec}
Haohe Liu, Xuenan Xu, Yi~Yuan, Mengyue Wu, Wenwu Wang, and Mark~D Plumbley,
\newblock ``Semanticodec: An ultra low bitrate semantic audio codec for general sound,''
\newblock {\em arXiv preprint arXiv:2405.00233}, 2024.

\bibitem{wang2023neural}
Chengyi Wang, Sanyuan Chen, Yu~Wu, Ziqiang Zhang, Long Zhou, Shujie Liu, Zhuo Chen, Yanqing Liu, Huaming Wang, Jinyu Li, et~al.,
\newblock ``Neural codec language models are zero-shot text to speech synthesizers,''
\newblock {\em arXiv preprint arXiv:2301.02111}, 2023.

\bibitem{kharitonov2022text}
Eugene Kharitonov, Ann Lee, Adam Polyak, Yossi Adi, Jade Copet, Kushal Lakhotia, Tu-Anh Nguyen, Morgane Rivi{\`e}re, Abdelrahman Mohamed, Emmanuel Dupoux, et~al.,
\newblock ``Text-free prosody-aware generative spoken language modeling,''
\newblock in {\em ACL 2022-Association for Computational Linguistics}. MIT Press, 2022, vol.~1, pp. 8666--8681.

\bibitem{rippel2014learning}
Oren Rippel, Michael Gelbart, and Ryan Adams,
\newblock ``Learning ordered representations with nested dropout,''
\newblock in {\em International Conference on Machine Learning}. PMLR, 2014, pp. 1746--1754.

\bibitem{xu2021anytime}
Yilun Xu, Yang Song, Sahaj Garg, Linyuan Gong, Rui Shu, Aditya Grover, and Stefano Ermon,
\newblock ``Anytime sampling for autoregressive models via ordered autoencoding,''
\newblock in {\em International Conference on Learning Representations}, 2021.

\bibitem{jegou2010product}
Herve Jegou, Matthijs Douze, and Cordelia Schmid,
\newblock ``{Product quantization for nearest neighbor search},''
\newblock {\em IEEE transactions on pattern analysis and machine intelligence}, vol. 33, no. 1, pp. 117--128, 2010.

\bibitem{guo2024addressing}
Haohan Guo, Fenglong Xie, Dongchao Yang, Hui Lu, Xixin Wu, and Helen Meng,
\newblock ``Addressing index collapse of large-codebook speech tokenizer with dual-decoding product-quantized variational auto-encoder,''
\newblock {\em arXiv preprint arXiv:2406.02940}, 2024.

\bibitem{dawalatabad2021ecapa}
Nauman Dawalatabad, Mirco Ravanelli, Fran{\c{c}}ois Grondin, Jenthe Thienpondt, Brecht Desplanques, and Hwidong Na,
\newblock ``Ecapa-tdnn embeddings for speaker diarization,''
\newblock {\em arXiv preprint arXiv:2104.01466}, 2021.

\bibitem{jiang2023mega}
Ziyue Jiang, Jinglin Liu, Yi~Ren, Jinzheng He, Chen Zhang, Zhenhui Ye, Pengfei Wei, Chunfeng Wang, Xiang Yin, Zejun Ma, et~al.,
\newblock ``Mega-tts 2: Zero-shot text-to-speech with arbitrary length speech prompts,''
\newblock {\em arXiv preprint arXiv:2307.07218}, 2023.

\bibitem{bigvgan}
{Sang-gil} Lee, Wei Ping, Boris Ginsburg, Bryan Catanzaro, and Sungroh Yoon,
\newblock ``Big{VGAN}: A universal neural vocoder with large-scale training,''
\newblock in {\em The Eleventh International Conference on Learning Representations}, 2023.

\bibitem{radford2019language}
Alec Radford, Jeffrey Wu, Rewon Child, David Luan, Dario Amodei, Ilya Sutskever, et~al.,
\newblock ``Language models are unsupervised multitask learners,''
\newblock {\em OpenAI blog}, vol. 1, no. 8, pp. 9, 2019.

\bibitem{ma2024wenetspeech4tts}
Linhan Ma, Dake Guo, Kun Song, Yuepeng Jiang, Shuai Wang, Liumeng Xue, Weiming Xu, Huan Zhao, Binbin Zhang, and Lei Xie,
\newblock ``Wenetspeech4tts: A 12,800-hour mandarin tts corpus for large speech generation model benchmark,'' 2024.

\bibitem{TencentGameMate}
Pengcheng Guo and Shixing Liu,
\newblock ``chinese speech pretrain,'' 2022.

\bibitem{loshchilov2017decoupled}
Ilya Loshchilov and Frank Hutter,
\newblock ``Decoupled weight decay regularization,''
\newblock {\em arXiv preprint arXiv:1711.05101}, 2017.

\bibitem{zhang2022wenetspeech}
Binbin Zhang, Hang Lv, Pengcheng Guo, Qijie Shao, Chao Yang, Lei Xie, Xin Xu, Hui Bu, Xiaoyu Chen, Chenchen Zeng, et~al.,
\newblock ``Wenetspeech: A 10000+ hours multi-domain mandarin corpus for speech recognition,''
\newblock in {\em ICASSP 2022-2022 IEEE International Conference on Acoustics, Speech and Signal Processing (ICASSP)}. IEEE, 2022, pp. 6182--6186.

\end{thebibliography}

\end{document}